\documentclass[twocolumn]{aastex62}
\usepackage{subfigure}
\usepackage{color}

\received{January 1, 2018}
\revised{January 7, 2018}
\accepted{\today}
\submitjournal{ApJ}

\shorttitle{TESS's first seismic known hosts}
\shortauthors{Campante et al.}

\begin{document}

\title{TESS ASTEROSEISMOLOGY OF THE KNOWN RED-GIANT HOST STARS HD~212771 AND HD~203949}

\correspondingauthor{Tiago L. Campante}
\email{tiago.campante@astro.up.pt}

\author[0000-0002-4588-5389]{Tiago L. Campante}
\affiliation{Instituto de Astrof\'{\i}sica e Ci\^{e}ncias do Espa\c{c}o, Universidade do Porto,  Rua das Estrelas, 4150-762 Porto, Portugal}
\affiliation{Departamento de F\'{\i}sica e Astronomia, Faculdade de Ci\^{e}ncias da Universidade do Porto, Rua do Campo Alegre, s/n, 4169-007 Porto, Portugal}
\affiliation{Kavli Institute for Theoretical Physics, University of California, Santa Barbara, CA 93106-4030, USA}

\author[0000-0001-8835-2075]{Enrico Corsaro}
\affiliation{INAF --- Osservatorio Astrofisico di Catania, via S.~Sofia 78, 95123 Catania, Italy}

\author[0000-0001-9214-5642]{Mikkel N. Lund}
\affiliation{Stellar Astrophysics Centre (SAC), Department of Physics and Astronomy, Aarhus University, Ny Munkegade 120, 8000 Aarhus C, Denmark}
\affiliation{Kavli Institute for Theoretical Physics, University of California, Santa Barbara, CA 93106-4030, USA}

\author[0000-0002-7547-1208]{Beno\^it Mosser}
\affiliation{LESIA, Observatoire de Paris, Universit\'e PSL, CNRS, Sorbonne Universit\'e, Universit\'e de Paris, 5 place Jules Janssen, 92195 Meudon, France}

\author[0000-0001-6359-2769]{Aldo Serenelli}
\affiliation{Institute of Space Sciences (ICE, CSIC) Campus UAB, Carrer de Can Magrans, s/n, E-08193, Bellaterra, Spain}
\affiliation{Institut d'Estudis Espacials de Catalunya (IEEC), C/Gran Capit\`a, 2-4, E-08034, Barcelona, Spain}
\affiliation{Kavli Institute for Theoretical Physics, University of California, Santa Barbara, CA 93106-4030, USA}

\author[0000-0001-8014-6162]{Dimitri Veras}
\altaffiliation{STFC Ernest Rutherford Fellow}
\affiliation{Centre for Exoplanets and Habitability, University of Warwick, Coventry CV4 7AL, UK}
\affiliation{Department of Physics, University of Warwick, Coventry CV4 7AL, UK}
\affiliation{Kavli Institute for Theoretical Physics, University of California, Santa Barbara, CA 93106-4030, USA}

\author[0000-0002-0601-6199]{Vardan Adibekyan}
\affiliation{Instituto de Astrof\'{\i}sica e Ci\^{e}ncias do Espa\c{c}o, Universidade do Porto,  Rua das Estrelas, 4150-762 Porto, Portugal}

\author[0000-0001-7549-9684]{H. M. Antia}
\affiliation{Tata Institute of Fundamental Research, Mumbai, India}

\author[0000-0002-4773-1017]{Warrick Ball}
\affiliation{School of Physics and Astronomy, University of Birmingham, Edgbaston, Birmingham B15 2TT, UK}
\affiliation{Stellar Astrophysics Centre (SAC), Department of Physics and Astronomy, Aarhus University, Ny Munkegade 120, 8000 Aarhus C, Denmark}

\author[0000-0002-6163-3472]{Sarbani Basu}
\affiliation{Department of Astronomy, Yale University, P.O.~Box 208101, New Haven, CT 06520-8101, USA}

\author[0000-0001-5222-4661]{Timothy R. Bedding}
\affiliation{Sydney Institute for Astronomy (SIfA), School of Physics, University of Sydney, NSW 2006, Australia}
\affiliation{Stellar Astrophysics Centre (SAC), Department of Physics and Astronomy, Aarhus University, Ny Munkegade 120, 8000 Aarhus C, Denmark}
\affiliation{Kavli Institute for Theoretical Physics, University of California, Santa Barbara, CA 93106-4030, USA}

\author[0000-0002-9480-8400]{Diego Bossini}
\affiliation{Instituto de Astrof\'{\i}sica e Ci\^{e}ncias do Espa\c{c}o, Universidade do Porto,  Rua das Estrelas, 4150-762 Porto, Portugal}

\author[0000-0002-4290-7351]{Guy R. Davies}
\affiliation{School of Physics and Astronomy, University of Birmingham, Edgbaston, Birmingham B15 2TT, UK}
\affiliation{Stellar Astrophysics Centre (SAC), Department of Physics and Astronomy, Aarhus University, Ny Munkegade 120, 8000 Aarhus C, Denmark}

\author[0000-0003-4434-2195]{Elisa Delgado Mena}
\affiliation{Instituto de Astrof\'{\i}sica e Ci\^{e}ncias do Espa\c{c}o, Universidade do Porto,  Rua das Estrelas, 4150-762 Porto, Portugal}

\author[0000-0002-8854-3776]{Rafael A. Garc\'ia}
\affiliation{IRFU, CEA, Universit\'e Paris-Saclay, F-91191 Gif-sur-Yvette, France}
\affiliation{AIM, CEA, CNRS, Universit\'e Paris-Saclay, Universit\'e Paris Diderot, Sorbonne Paris Cit\'e, F-91191 Gif-sur-Yvette, France}

\author[0000-0001-8725-4502]{Rasmus Handberg}
\affiliation{Stellar Astrophysics Centre (SAC), Department of Physics and Astronomy, Aarhus University, Ny Munkegade 120, 8000 Aarhus C, Denmark}

\author[0000-0003-2400-6960]{Marc Hon}
\affiliation{School of Physics, The University of New South Wales, Sydney NSW 2052, Australia}

\author[0000-0002-7084-0529]{Stephen R. Kane}
\affiliation{Department of Earth and Planetary Sciences, University of California, Riverside, CA 92521, USA}

\author[0000-0002-6536-6367]{Steven D. Kawaler}
\affiliation{Department of Physics and Astronomy, Iowa State University, Ames, IA 50011, USA}
\affiliation{Kavli Institute for Theoretical Physics, University of California, Santa Barbara, CA 93106-4030, USA}

\author[0000-0002-3322-5279]{James S. Kuszlewicz}
\affiliation{Max-Planck-Institut f\"ur Sonnensystemforschung, Justus-von-Liebig-Weg 3, 37077 G\"ottingen, Germany}
\affiliation{Stellar Astrophysics Centre (SAC), Department of Physics and Astronomy, Aarhus University, Ny Munkegade 120, 8000 Aarhus C, Denmark}

\author{Miles Lucas}
\affiliation{Department of Physics and Astronomy, Iowa State University, Ames, IA 50011, USA}

\author[0000-0002-0129-0316]{Savita Mathur}
\affiliation{Instituto de Astrof\'isica de Canarias (IAC), E-38205 La Laguna, Tenerife, Spain}
\affiliation{Universidad de La Laguna (ULL), Departamento de Astrof\'isica, E-38206 La Laguna, Tenerife, Spain}

\author[0000-0002-7399-0231]{Nicolas Nardetto}
\affiliation{Universit\'e C\^{o}te d'Azur, Observatoire de la C\^{o}te d'Azur, CNRS, Laboratoire Lagrange, France}

\author[0000-0001-9169-2599]{Martin B. Nielsen}
\affiliation{School of Physics and Astronomy, University of Birmingham, Edgbaston, Birmingham B15 2TT, UK}
\affiliation{Stellar Astrophysics Centre (SAC), Department of Physics and Astronomy, Aarhus University, Ny Munkegade 120, 8000 Aarhus C, Denmark}
\affiliation{Center for Space Science, NYUAD Institute, New York University Abu Dhabi, PO Box 129188, Abu Dhabi, UAE}

\author[0000-0002-7549-7766]{Marc H. Pinsonneault}
\affiliation{Department of Astronomy, The Ohio State University, Columbus, OH 43210, USA}
\affiliation{Kavli Institute for Theoretical Physics, University of California, Santa Barbara, CA 93106-4030, USA}

\author[0000-0002-0460-8289]{Sabine Reffert}
\affiliation{Landessternwarte, Zentrum f\"ur Astronomie der Universit\"at Heidelberg, K\"onigstuhl 12, 69117 Heidelberg, Germany}

\author[0000-0002-6137-903X]{V\'ictor Silva Aguirre}
\affiliation{Stellar Astrophysics Centre (SAC), Department of Physics and Astronomy, Aarhus University, Ny Munkegade 120, 8000 Aarhus C, Denmark}

\author[0000-0002-3481-9052]{Keivan G. Stassun}
\affiliation{Vanderbilt University, Department of Physics and Astronomy, 6301 Stevenson Center Ln., Nashville, TN 37235, USA}
\affiliation{Vanderbilt Initiative in Data-intensive Astrophysics (VIDA), 6301 Stevenson Center Ln., Nashville, TN 37235, USA}

\author[0000-0002-4879-3519]{Dennis Stello}
\affiliation{School of Physics, The University of New South Wales, Sydney NSW 2052, Australia}
\affiliation{Sydney Institute for Astronomy (SIfA), School of Physics, University of Sydney, NSW 2006, Australia}
\affiliation{Stellar Astrophysics Centre (SAC), Department of Physics and Astronomy, Aarhus University, Ny Munkegade 120, 8000 Aarhus C, Denmark}
\affiliation{Kavli Institute for Theoretical Physics, University of California, Santa Barbara, CA 93106-4030, USA}

\author[0000-0002-1166-9338]{Stephan Stock}
\affiliation{Landessternwarte, Zentrum f\"ur Astronomie der Universit\"at Heidelberg, K\"onigstuhl 12, 69117 Heidelberg, Germany}

\author{Mathieu Vrard}
\affiliation{Instituto de Astrof\'{\i}sica e Ci\^{e}ncias do Espa\c{c}o, Universidade do Porto,  Rua das Estrelas, 4150-762 Porto, Portugal}

\author[0000-0002-7772-7641]{Mutlu Y{\i}ld{\i}z}
\affiliation{Department of Astronomy and Space Sciences, Science Faculty, Ege University, 35100, Bornova, \.{I}zmir, Turkey}

\author[0000-0002-5714-8618]{William J. Chaplin}
\affiliation{School of Physics and Astronomy, University of Birmingham, Edgbaston, Birmingham B15 2TT, UK}
\affiliation{Stellar Astrophysics Centre (SAC), Department of Physics and Astronomy, Aarhus University, Ny Munkegade 120, 8000 Aarhus C, Denmark}
\affiliation{Kavli Institute for Theoretical Physics, University of California, Santa Barbara, CA 93106-4030, USA}

\author[0000-0001-8832-4488]{Daniel Huber}
\affiliation{Institute for Astronomy, University of Hawai`i, 2680 Woodlawn Drive, Honolulu, HI 96822, USA}
\affiliation{Kavli Institute for Theoretical Physics, University of California, Santa Barbara, CA 93106-4030, USA}

\author[0000-0003-4733-6532]{Jacob L. Bean}
\affiliation{Department of Astronomy and Astrophysics, University of Chicago, 5640 S.~Ellis Avenue, Chicago, IL 60637, USA}

\author[0000-0002-9424-2339]{Zeynep \c{C}el\.ik Orhan}
\affiliation{Department of Astronomy and Space Sciences, Science Faculty, Ege University, 35100, Bornova, \.{I}zmir, Turkey}

\author[0000-0001-8237-7343]{Margarida S. Cunha}
\affiliation{Instituto de Astrof\'{\i}sica e Ci\^{e}ncias do Espa\c{c}o, Universidade do Porto,  Rua das Estrelas, 4150-762 Porto, Portugal}
\affiliation{Departamento de F\'{\i}sica e Astronomia, Faculdade de Ci\^{e}ncias da Universidade do Porto, Rua do Campo Alegre, s/n, 4169-007 Porto, Portugal}

\author[0000-0001-5137-0966]{J\o rgen Christensen-Dalsgaard}
\affiliation{Stellar Astrophysics Centre (SAC), Department of Physics and Astronomy, Aarhus University, Ny Munkegade 120, 8000 Aarhus C, Denmark}
\affiliation{Kavli Institute for Theoretical Physics, University of California, Santa Barbara, CA 93106-4030, USA}

\author[0000-0002-9037-0018]{Hans Kjeldsen}
\affiliation{Stellar Astrophysics Centre (SAC), Department of Physics and Astronomy, Aarhus University, Ny Munkegade 120, 8000 Aarhus C, Denmark}
\affiliation{Institute of Theoretical Physics and Astronomy, Vilnius University, Saul\.{e}tekio av.~3, 10257 Vilnius, Lithuania}

\author[0000-0003-4034-0416]{Travis S. Metcalfe}
\affiliation{Space Science Institute, 4750 Walnut Street, Suite 205, Boulder, CO 80301, USA}
\affiliation{Max-Planck-Institut f\"ur Sonnensystemforschung, Justus-von-Liebig-Weg 3, 37077 G\"ottingen, Germany}

\author[0000-0001-5998-8533]{Andrea Miglio}
\affiliation{School of Physics and Astronomy, University of Birmingham, Edgbaston, Birmingham B15 2TT, UK}
\affiliation{Stellar Astrophysics Centre (SAC), Department of Physics and Astronomy, Aarhus University, Ny Munkegade 120, 8000 Aarhus C, Denmark}

\author[0000-0003-0513-8116]{M\'ario J. P. F. G. Monteiro}
\affiliation{Instituto de Astrof\'{\i}sica e Ci\^{e}ncias do Espa\c{c}o, Universidade do Porto,  Rua das Estrelas, 4150-762 Porto, Portugal}
\affiliation{Departamento de F\'{\i}sica e Astronomia, Faculdade de Ci\^{e}ncias da Universidade do Porto, Rua do Campo Alegre, s/n, 4169-007 Porto, Portugal}

\author[0000-0002-4647-2068]{Benard Nsamba}
\affiliation{Instituto de Astrof\'{\i}sica e Ci\^{e}ncias do Espa\c{c}o, Universidade do Porto,  Rua das Estrelas, 4150-762 Porto, Portugal}

\author[0000-0001-5759-7790]{S\.ibel \"Ortel}
\affiliation{Department of Astronomy and Space Sciences, Science Faculty, Ege University, 35100, Bornova, \.{I}zmir, Turkey}

\author[0000-0002-2157-7146]{Filipe Pereira}
\affiliation{Instituto de Astrof\'{\i}sica e Ci\^{e}ncias do Espa\c{c}o, Universidade do Porto,  Rua das Estrelas, 4150-762 Porto, Portugal}

\author[0000-0001-9047-2965]{S\'ergio G. Sousa}
\affiliation{Instituto de Astrof\'{\i}sica e Ci\^{e}ncias do Espa\c{c}o, Universidade do Porto,  Rua das Estrelas, 4150-762 Porto, Portugal}
\affiliation{Departamento de F\'{\i}sica e Astronomia, Faculdade de Ci\^{e}ncias da Universidade do Porto, Rua do Campo Alegre, s/n, 4169-007 Porto, Portugal}

\author[0000-0002-0552-2313]{Maria Tsantaki}
\affiliation{Instituto de Astrof\'{\i}sica e Ci\^{e}ncias do Espa\c{c}o, Universidade do Porto,  Rua das Estrelas, 4150-762 Porto, Portugal}

\author{Margaret C. Turnbull}
\affiliation{SETI Institute, Carl Sagan Center for the Study of Life in the Universe, Off-Site: 2801 Shefford Drive, Madison, WI 53719, USA}



\begin{abstract}

The \textit{Transiting Exoplanet Survey Satellite} (TESS) is performing a near all-sky survey for planets that transit bright stars. In addition, its excellent photometric precision enables asteroseismology of solar-type and red-giant stars, which exhibit convection-driven, solar-like oscillations. Simulations predict that TESS will detect solar-like oscillations in nearly 100 stars already known to host planets. In this paper, we present an asteroseismic analysis of the known red-giant host stars HD~212771 and HD~203949, both systems having a long-period planet detected through radial velocities. These are the first detections of oscillations in previously known exoplanet-host stars by TESS, further showcasing the mission's potential to conduct asteroseismology of red-giant stars. We estimate the fundamental properties of both stars through a grid-based modeling approach that uses global asteroseismic parameters as input. We discuss the evolutionary state of HD~203949 in depth and note the large discrepancy between its asteroseismic mass ($M_\ast = 1.23 \pm 0.15\,{\rm M}_\odot$ if on the red-giant branch or $M_\ast = 1.00 \pm 0.16\,{\rm M}_\odot$ if in the clump) and the mass quoted in the discovery paper ($M_\ast = 2.1 \pm 0.1\,{\rm M}_\odot$), implying a change $>30\,\%$ in the planet's mass. Assuming HD~203949 to be in the clump, we investigate the planet's past orbital evolution and discuss how it could have avoided engulfment at the tip of the red-giant branch. Finally, HD~212771 was observed by K2 during its Campaign 3, thus allowing for a preliminary comparison of the asteroseismic performances of TESS and K2. We estimate the ratio of the observed oscillation amplitudes for this star to be $A_{\rm max}^{\rm TESS}/A_{\rm max}^{\rm K2} = 0.75 \pm 0.14$, consistent with the expected ratio of $\sim0.85$ due to the redder bandpass of TESS.

\end{abstract}



\section{Introduction} \label{sec:intro}

Major advances in stellar interiors physics and evolution have recently been made possible by asteroseismology. This has largely been due to the exquisite space-based data made available by CNES/ESA's CoRoT \citep{Baglin09} and NASA's \textit{Kepler}/K2 \citep{Borucki10,Koch10,Howell14} missions. In particular, asteroseismology has vastly benefited the study of solar-type and red-giant stars, which exhibit convection-driven, solar-like oscillations \citep[for a review, see][]{ChaplinMiglio13}. The revolution triggered by CoRoT and \textit{Kepler}/K2 is set to continue over the coming decade, with
NASA's TESS \citep{Ricker14} and ESA's PLATO \citep{Rauer14} missions expected to raise the number of
known solar-like oscillators by up to two orders of magnitude \citep{Huber18}.

Fueled by the wealth of high-quality seismic data, the past few years have witnessed an ever-growing effort being devoted to the development of novel techniques for the estimation of fundamental stellar properties. The focus has been placed on uniform data analysis \citep[e.g.,][]{Davies16,Lund17} and stellar modeling \citep[e.g.,][]{Serenelli17,VSA17,Nsamba18} strategies, as well as on state-of-the-art optimization procedures that make use of individual oscillation frequencies \citep[e.g.,][]{Metcalfe10,Mathur12,VSA15,Rendle19}.

These techniques make it possible to estimate precise properties of large numbers of field stars, for which such information is sparse. As a result, asteroseismology is having a profound impact on modern astrophysics, notably on the field of exoplanetary science \citep{CampanteBook}. Characterization of exoplanet-host stars via asteroseismology allows for unmatched precision in the absolute properties of their planets \citep{Huber13,Ballard14,Kepler444,VSA15,Lundkvist16}. Furthermore, asteroseismology enables constraints on the spin-orbit alignment of exoplanet systems \citep{Chaplin13,HuberSci,CampanteAlign,Kamiaka19} as well as statistical inferences on orbital eccentricities via asterodensity profiling \citep{SliskiKipping14,VanEylenAlbrecht15,VanEylen19}.

The \textit{Transiting Exoplanet Survey Satellite} (TESS) is performing a near all-sky survey for planets that transit bright stars. Moreover, its excellent photometric precision, combined with its fine time sampling and long intervals of uninterrupted observations, enables asteroseismology of solar-like oscillators \citep{Campante16,Schofield19}. In particular, simulations predict that TESS will detect solar-like oscillations in nearly 100 solar-type and red-giant stars already known to host planets \citep{Campante16}.

In this paper, we present an asteroseismic analysis of the evolved known hosts HD~212771 and HD~203949, both systems having a long-period planet detected through the radial-velocity (RV) method. These are the first detections of oscillations in previously known exoplanet-host stars by TESS and follow the discovery of the first planet transiting a star in which oscillations could be measured \citep[TOI-197 or TESS Object of Interest 197;][]{TOI197}.

HD~212771 (TIC~12723961, HIP~110813) is a bright (with apparent TESS magnitude $T=6.75$), spectroscopically-classified subgiant \citep[G8~IV;][]{HoukSmith-Moore88}, being among the targets of the RV planet survey of \citet{Johnson07}. It hosts a Jovian planet with minimum mass $M_{\rm p}\,\sin i = 2.3 \pm 0.4\,M_{\rm J}$ in a 373.3-day orbit \citep{Johnson10}. HD~212771 was subsequently observed by K2 in short cadence during its Campaign 3, spanning a total of $\sim\,$69 days. This allowed estimation of its fundamental properties through a grid-based modeling approach that used global asteroseismic parameters, complementary spectroscopy and a parallax-based luminosity as input \citep{Campante17,North17}.

HD~203949 (TIC~129649472, HIP~105854) is a bright ($T=4.75$), spectroscopically-classified giant \citep[K2~III;][]{Houk82}. A massive planet ($M_{\rm p}\,\sin i = 8.2 \pm 0.2\,M_{\rm J}$) was discovered in a 184.2-day circular ($e=0.02\pm0.03$) orbit around HD~203949 by \citet{Jones14} as part of the EXoPlanets aRound Evolved StarS (EXPRESS) project \citep{Jones11}.

The rest of this paper is organized as follows. In Sect.~\ref{sec:obs}, we present the available observational data (including the TESS photometry). This is followed by an asteroseismic analysis (Sect.~\ref{sec:seismo}) and the estimation of fundamental stellar properties through a grid-based modeling approach (Sect.~\ref{sec:properties}). Finally, we discuss our results in Sect.~\ref{sec:discuss} and provide an outlook in Sect.~\ref{sec:conclusions}.

\section{Observations}\label{sec:obs}

\subsection{TESS Photometry}\label{sec:TESSphot}

TESS observed HD~212771 and HD~203949 in 2-minute cadence over 27.4 days during Sectors 2 and 1 of Cycle 1, respectively. Both targets were part of a larger cohort of 79 ``fast-track'' targets that were processed using a special version \citep{TDAdr} of the TESS Asteroseismic Science Operations Center\footnote{\url{https://tasoc.dk/}} \citep[TASOC;][]{TASOC} photometry pipeline\footnote{\url{https://github.com/tasoc}}. Starting from calibrated target pixel files, aperture photometry was conducted following a procedure similar to the one adopted in the \texttt{K2P$^{2}$} pipeline \citep{K2P2}, originally developed to generate light curves from data collected by K2. The extracted light curves were subsequently corrected for systematic effects using the KASOC filter \citep{KASOCfilter}.

\begin{figure*}[!t]
\centering
  \subfigure{%
  \includegraphics[width=.48\textwidth,trim={2cm 8.5cm 1cm 8.75cm},clip]{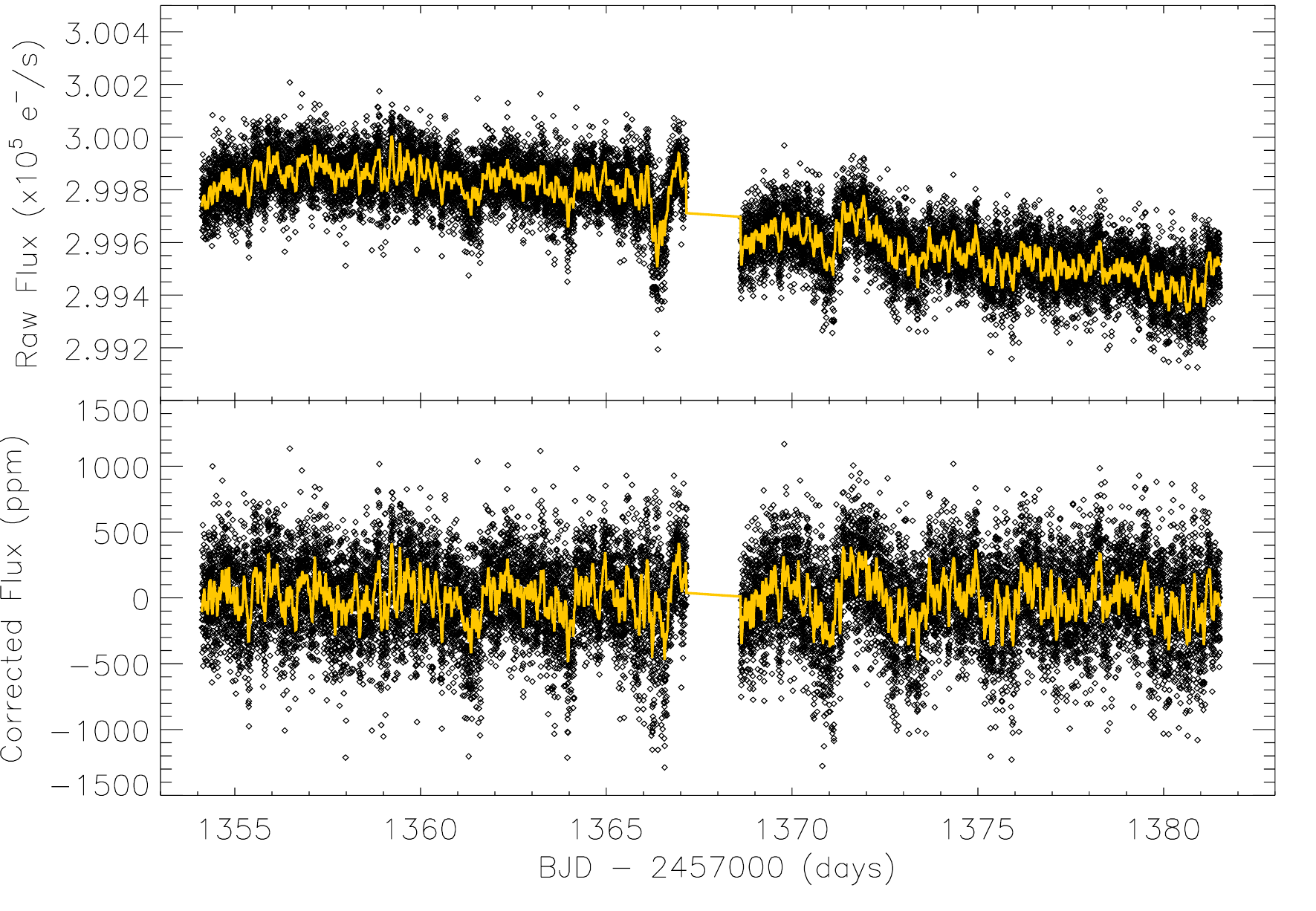}}\hfill
  \subfigure{%
  \includegraphics[width=.48\textwidth,trim={2cm 8.5cm 1cm 8.75cm},clip]{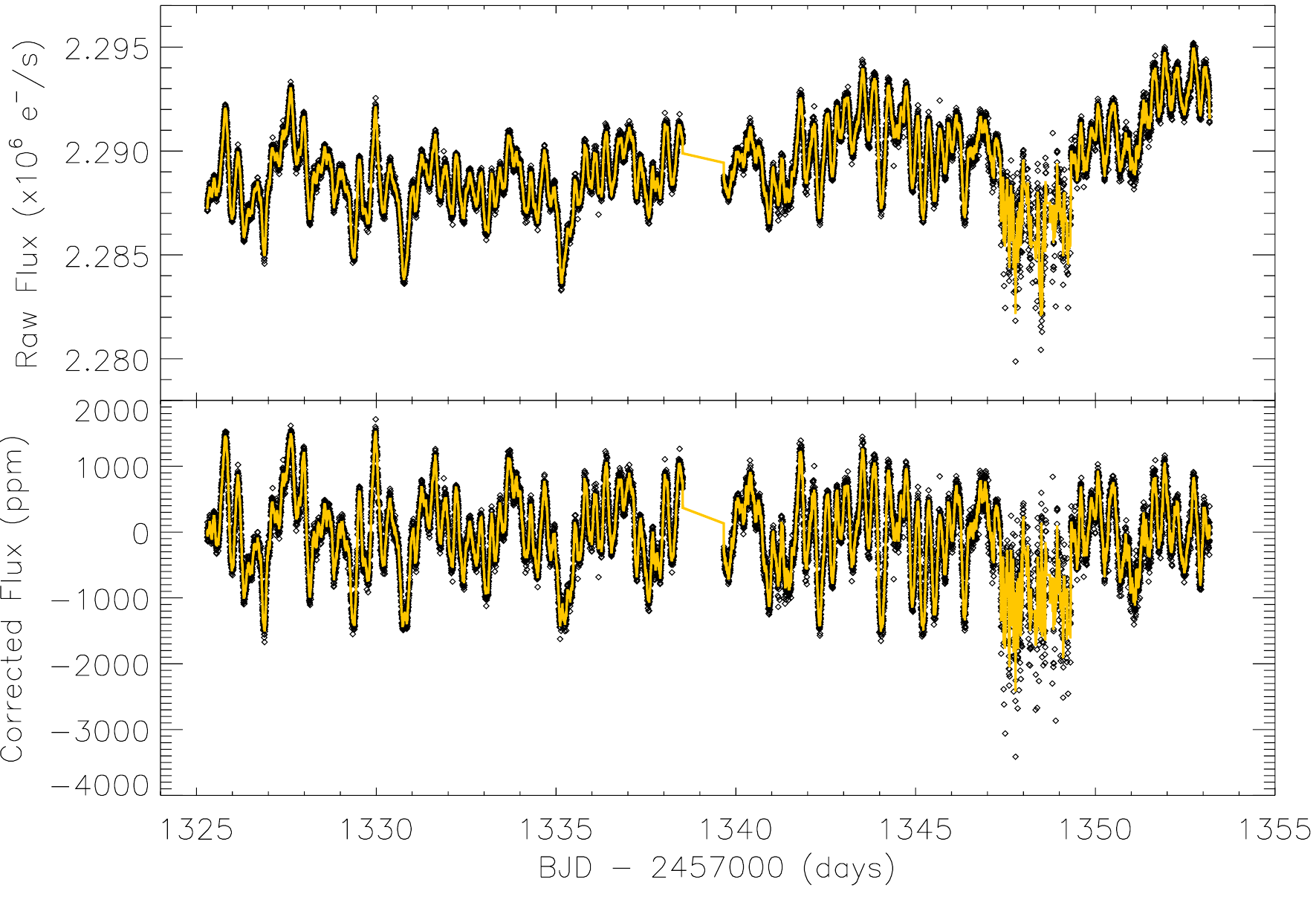}}
  \caption{Light curves of HD~212771 (left panel) and HD~203949 (right panel) produced by the TASOC photometry pipeline. In each panel, raw (top) and corrected (bottom) 2-minute cadence light curves are displayed. A smoothed --- using a 1-hour (HD~212771) and 10-minute (HD~203949) boxcar filters --- version of the light curve is depicted by a yellow curve in each subpanel.}\label{fig:LCs}
\end{figure*}

Figure \ref{fig:LCs} shows the light curves of HD~212771 (left panel) and HD~203949 (right panel) produced by the TASOC pipeline. Both light curves have high duty cycles ($\sim$98\,\% and $\sim$94\,\%, respectively), displaying a gap midway through (due to the data downlink) that separates the two spacecraft orbits in each sector. A 2.5-day periodicity can be seen, especially in the bottom left subpanel, caused by the spacecraft's angular momentum dumping cycle. Moreover, a region of large jitter can be seen in the right panel towards the end of the sector, a feature common to Sector 1 pointings \citep{TDAdr}.

\subsection{High-Resolution Spectroscopy}\label{sec:spec}

We adopt the atmospheric parameters and elemental abundances obtained for HD~212771 by \citet{Campante17}, which are based on the analysis of a high-quality FEROS spectrum (see Table \ref{tab:HD212771}).

Particular care must, however, be taken regarding HD~203949, since use of the model-independent scaling relation $g \propto \nu_{\rm max}\,T_{\rm eff}^{1/2}$ \citep{Brown91,KB95,Belkacem11}, where $\nu_{\rm max}$ is the frequency of maximum oscillation amplitude (see Sect.~\ref{sec:globalseismo}), leads to an initial estimate of the surface gravity ($\log g \sim 2.4$) that is significantly lower than the spectroscopic value quoted in the discovery paper \citep[$\log g = 2.94 \pm 0.20$;][]{Jones11,Jones14}.

Therefore, we instead adopt the spectroscopic parameters listed for HD~203949 in the SWEET-Cat catalog\footnote{\url{https://www.astro.up.pt/resources/sweet-cat/}} \citep{SWEET}, whose $\log g$ is compatible with the one inferred from asteroseismology. These are based on the analysis of a high-resolution ($R \sim 100{,}000$) UVES spectrum \citep[for details, see][]{Sousa18} and are listed in Table \ref{tab:HD203949}. Analysis of a lower-resolution ($R \sim 48{,}000$) FEROS spectrum by the same authors, albeit considering a larger number of iron lines, led to fully consistent spectroscopic parameters.

We conducted a detailed abundance analysis of HD~203949 based on both the UVES and FEROS spectra, following the methodology described by \citet{Adibekyan-12,Adibekyan-15a}. Abundances derived from the two spectra are consistent within $1\sigma$, with UVES-based results being slightly more precise. Our analysis shows that HD~203949 has a chemical composition typical of a thin-disk, metal-rich K giant \citep[e.g.,][]{Adibekyan-15b,Jofre-15}. The star shows enhancement in Na and Al relative to iron ([Na/Fe] = 0.22$\,\pm\,$0.13 dex and [Al/Fe] = 0.25$\,\pm\,$0.12 dex based on the UVES spectrum), typical of evolved stars \citep[e.g.,][]{Adibekyan-15b}. Finally, we computed the relative abundance of $\alpha$ elements as the unweighted mean of the Mg, Si, Ca and Ti abundances derived from the UVES spectrum, resulting in $[\alpha/{\rm Fe}] = 0.07 \pm 0.09\:{\rm dex}$.

\subsection{Broadband Photometric Spectral Energy Distribution}\label{sec:sed}

We fitted the spectral energy distributions (SEDs) of both stars using broadband photometry to determine empirical constraints on the stellar radii and luminosities, following the method described in \citet{StassunTorres16} and \citet{Stassun2018}. The available broadband photometry in published all-sky catalogs (i.e., APASS, 2MASS and WISE) provides coverage over the wavelength range $\approx\,$0.4--$22\:{\rm \mu m}$. For each star, we fitted a standard \citet{Kurucz2013} stellar atmosphere model, selected according to the spectroscopically-determined $T_{\rm eff}$, $\log g$ and [Fe/H] (see Sect.~\ref{sec:spec}). With these constraints fixed, the remaining free parameter in the fit was the extinction, $A_V$, which we limited to the maximum for the line of sight from the dust maps of \citet{Schlegel1998}. Finally, we integrated the (non-reddened) SED to obtain the bolometric flux at Earth ($F_{\rm bol}$) which, with the $T_{\rm eff}$ and the \textit{Gaia} DR2 distance \citep[adjusted for the systematic offset of][]{StassunTorres2018}, gives the stellar radius.

The best-fit parameters for HD~212771, with reduced $\chi^2 = 4.7$, are: $A_V = 0.04 \pm 0.04$, $F_{\rm bol} = (3.06\pm 0.14) \times 10^{-8}$ erg~s$^{-1}$~cm$^{-2}$, resulting in $R_\star = 4.44 \pm 0.13$ R$_\odot$ and $L_\star = 11.67 \pm 0.57$ L$_\odot$.
For HD~203949, the best-fit parameters, with reduced $\chi^2 = 3.7$, are: $A_V = 0.13 \pm 0.10$, $F_{\rm bol} = (2.27 \pm 0.22) \times 10^{-7}$ erg~s$^{-1}$~cm$^{-2}$, resulting in $R_\star = 10.30 \pm 0.51$ R$_\odot$ and $L_\star = 43.34 \pm 4.27$ L$_\odot$. The derived luminosities will be used in Sect.~\ref{sec:properties} as input to the grid-based modeling.

\subsection{Surface Brightness-Color Relation}\label{sec:sbcr}

\citet{Pietrzynski19} derived the distance to the Large Magellanic Cloud with a 1\,\% precision using eclipsing binaries as distance indicators. In order to achieve such precision, they used a dedicated surface brightness-color relation (SBCR) based on the observation of 48 red-clump stars with the PIONIER/VLTI instrument \citep{Gallenne18}. We used this relation to place empirical constraints on the angular diameters and linear radii of HD~212771 and HD~203949.

For HD~212771, considering $V = 7.60 \pm 0.01$ \citep{Hog00}, $K = 5.50 \pm 0.02$ \citep{Cutri03}, $A_V = 0.005$ \citep[using the Stilism dust map;][]{Lallement14,Capitanio17}, and $A_K = 0.089 A_V$ \citep{Nishiyama09}, we obtained \citep[using eq.~2 of][]{Pietrzynski19} a limb-darkened angular diameter of $\theta = 0.375 \pm 0.003\:{\rm mas}$. The quoted uncertainty (0.8\,\%) arises from the rms scatter (0.018 mag) of the SBCR in \citet{Pietrzynski19}. Using the \textit{Gaia} DR2 parallax, we then derived a stellar radius of $R_\ast = 4.45 \pm 0.04\,{\rm R}_\odot$. We must, however, also consider the source of uncertainty associated with the 2MASS infrared photometry (0.02 mag), which corresponds to an additional uncertainty in the radius of $0.05\,{\rm R}_\odot$. We thus finally obtained $R_\ast = 4.45 \pm 0.07\,{\rm R}_\odot$. Applying the same procedure to HD~203949 ($V = 5.62 \pm 0.01$, $K = 2.99 \pm 0.24$, and $A_V = 0.004$), we obtained a limb-darkened angular diameter of $\theta = 1.284 \pm 0.011\:{\rm mas}$ and a stellar radius of $R_\ast = 10.8 \pm 1.6\,{\rm R}_\odot$ (after taking into account the exceptionally large uncertainty of 0.24 mag in the infrared photometry). The derived stellar radii are consistent with those obtained from the SED analysis.


\begin{table}[!t]
\begin{center}
\caption{Stellar Parameters for HD~212771}
\renewcommand{\tabcolsep}{3mm}
\begin{tabular}{l c c}
\noalign{\smallskip}
\tableline\tableline
\noalign{\smallskip}
\textbf{Parameter} & \textbf{Value} & \textbf{Source} \\
\noalign{\smallskip}
\tableline
\noalign{\smallskip}
\multicolumn{3}{c}{Basic Properties} \\
\noalign{\smallskip}
\hline
\noalign{\smallskip}
TIC & 12723961 & 1 \\
\textit{Hipparcos} ID & 110813 & 2 \\
TESS Mag. & 6.75 & 1 \\
Sp.~Type & G8 IV & 3 \\
\noalign{\smallskip}
\hline
\noalign{\smallskip}
\multicolumn{3}{c}{Spectroscopy} \\
\noalign{\smallskip}
\hline
\noalign{\smallskip}
$T_{\rm eff}$ (K) & $5065 \pm 75$ & 4 \\
$[{\rm Fe}/{\rm H}]$ (dex) & $-0.10 \pm 0.10$ & 4 \\
$[\alpha/{\rm Fe}]$ (dex) & $0.06 \pm 0.05$\tablenotemark{a} & 4 \\
$\log g$ (cgs) & $3.37 \pm 0.17$ & 4 \\
\noalign{\smallskip}
\hline
\noalign{\smallskip}
\multicolumn{3}{c}{SED \& \textit{Gaia} DR2 Parallax} \\
\noalign{\smallskip}
\hline
\noalign{\smallskip}
$A_V$ & $0.04 \pm 0.04$ & 5 \\
$F_{\rm bol}$ (${\rm erg\,s^{-1}\,cm^{-2}}$) & $(3.06\pm 0.14) \times 10^{-8}$ & 5 \\
$R_\ast$ (${\rm R}_\odot$) & $4.44 \pm 0.13$ & 5 \\
$L_\ast$ (${\rm L}_\odot$) & $11.67 \pm 0.57$ & 5 \\
$\pi$ (mas) & $9.050 \pm 0.055$\tablenotemark{b} & 6 \\
\noalign{\smallskip}
\hline
\noalign{\smallskip}
\multicolumn{3}{c}{SBCR} \\
\noalign{\smallskip}
\hline
\noalign{\smallskip}
$\theta$ (mas) & $0.375 \pm 0.003$ & 5 \\
$R_\ast$ (${\rm R}_\odot$) & $4.45 \pm 0.07$ & 5 \\
\noalign{\smallskip}
\hline
\noalign{\smallskip}
\multicolumn{3}{c}{Asteroseismology} \\
\noalign{\smallskip}
\hline
\noalign{\smallskip}
$\Delta\nu$ ($\mu$Hz) & $16.25 \pm 0.19$ & 5 \\
$\nu_{\rm max}$ ($\mu$Hz) & $226.6 \pm 9.4$ & 5 \\
$\Delta\Pi_1$ (s) & $85.3 \pm 0.3$ & 5 \\
$M_\ast$ (${\rm M}_\odot$) & $1.42 \pm 0.07$ & 5 \\
$R_\ast$ (${\rm R}_\odot$) & $4.61 \pm 0.09$ & 5 \\
$\rho_\ast$ (gcc) & $0.02048 \pm 0.00050$ & 5 \\
$\log g$ (cgs) & $3.263 \pm 0.010$ & 5 \\
$t$ (Gyr) & $2.90 \pm 0.47$ & 5 \\
\noalign{\smallskip}
\hline
\noalign{\smallskip}
\end{tabular}
\end{center}
\tablenotetext{a}{\scriptsize The uncertainty (0.02) reported in (4) is not correct.}
\tablenotetext{b}{\scriptsize Adjusted for the systematic offset of \citet{StassunTorres2018}.}
\tablerefs{\scriptsize (1) \citet{TIC}, (2) \citet{HIP}, (3) \citet{HoukSmith-Moore88}, (4) \citet{Campante17}, (5) this work, (6) \citet{GaiaDR2}.}
\label{tab:HD212771}
\end{table}

\begin{table}[!t]
\begin{center}
\caption{Stellar Parameters for HD~203949}
\renewcommand{\tabcolsep}{0mm}
\begin{tabular}{l c c}
\noalign{\smallskip}
\tableline\tableline
\noalign{\smallskip}
\textbf{Parameter} & \textbf{Value} & \textbf{Source} \\
\noalign{\smallskip}
\tableline
\noalign{\smallskip}
\multicolumn{3}{c}{Basic Properties} \\
\noalign{\smallskip}
\hline
\noalign{\smallskip}
TIC & 129649472 & 1 \\
\textit{Hipparcos} ID & 105854 & 2 \\
TESS Mag. & 4.75 & 1 \\
Sp.~Type & K2 III & 3 \\
\noalign{\smallskip}
\hline
\noalign{\smallskip}
\multicolumn{3}{c}{Spectroscopy} \\
\noalign{\smallskip}
\hline
\noalign{\smallskip}
$T_{\rm eff}$ (K) & $4618 \pm 113$ & 4 \\
$[{\rm Fe}/{\rm H}]$ (dex) & $0.17 \pm 0.07$ & 4 \\
$[\alpha/{\rm Fe}]$ (dex) & $0.07 \pm 0.09$ & 5 \\
$\log g$ (cgs) & $2.36 \pm 0.28$ & 4 \\
\noalign{\smallskip}
\hline
\noalign{\smallskip}
\multicolumn{3}{c}{SED \& \textit{Gaia} DR2 Parallax} \\
\noalign{\smallskip}
\hline
\noalign{\smallskip}
$A_V$ & $0.13 \pm 0.10$ & 5 \\
$F_{\rm bol}$ (${\rm erg\,s^{-1}\,cm^{-2}}$) & $(2.27 \pm 0.22) \times 10^{-7}$ & 5 \\
$R_\ast$ (${\rm R}_\odot$) & $10.30 \pm 0.51$ & 5 \\
$L_\ast$ (${\rm L}_\odot$) & $43.34 \pm 4.27$ & 5 \\
$\pi$ (mas) & $12.77 \pm 0.13$\tablenotemark{a} & 6 \\
\noalign{\smallskip}
\hline
\noalign{\smallskip}
\multicolumn{3}{c}{SBCR} \\
\noalign{\smallskip}
\hline
\noalign{\smallskip}
$\theta$ (mas) & $1.284 \pm 0.011$ & 5 \\
$R_\ast$ (${\rm R}_\odot$) & $10.8 \pm 1.6$ & 5 \\
\noalign{\smallskip}
\hline
\noalign{\smallskip}
\multicolumn{3}{c}{Asteroseismology\tablenotemark{b}} \\
\noalign{\smallskip}
\hline
\noalign{\smallskip}
$\Delta\nu$ ($\mu$Hz) & $4.10 \pm 0.14$ & 5 \\
$\nu_{\rm max}$ ($\mu$Hz) & $31.6 \pm 3.2$ & 5 \\
$\Delta\Pi_1$ (s) & \nodata & \nodata \\
$M_\ast$ (${\rm M}_\odot$) & $1.23 \pm 0.15$ / $1.00 \pm 0.16$ & 5 \\
$R_\ast$ (${\rm R}_\odot$) & $10.93 \pm 0.54$ / $10.34 \pm 0.55$ & 5 \\
$\rho_\ast$ (gcc) & $0.00134 \pm 0.00010$ / $0.00130 \pm 0.00011$ & 5 \\
$\log g$ (cgs) & $2.453 \pm 0.027$ / $2.415 \pm 0.044$ & 5 \\
$t$ (Gyr) & $6.45 \pm 2.79$ / $7.29 \pm 3.06$ & 5 \\
\noalign{\smallskip}
\hline
\noalign{\smallskip}
\end{tabular}
\end{center}
\tablenotetext{a}{\scriptsize Adjusted for the systematic offset of \citet{StassunTorres2018}.}
\tablenotetext{b}{\scriptsize Fundamental stellar properties are provided assuming that the star is either on the RGB or in the clump, respectively (see Sect.~\ref{sec:evolstate}).}
\tablerefs{\scriptsize (1) \citet{TIC}, (2) \citet{HIP}, (3) \citet{Houk82}, (4) \citet{Sousa18}, (5) this work, (6) \citet{GaiaDR2}.}
\label{tab:HD203949}
\end{table}

\section{Asteroseismology}\label{sec:seismo}

\begin{figure*}[!t]
\centering
  \subfigure{%
  \includegraphics[width=.48\textwidth,trim={0.5cm 1.75cm 0.5cm 1.5cm},clip]{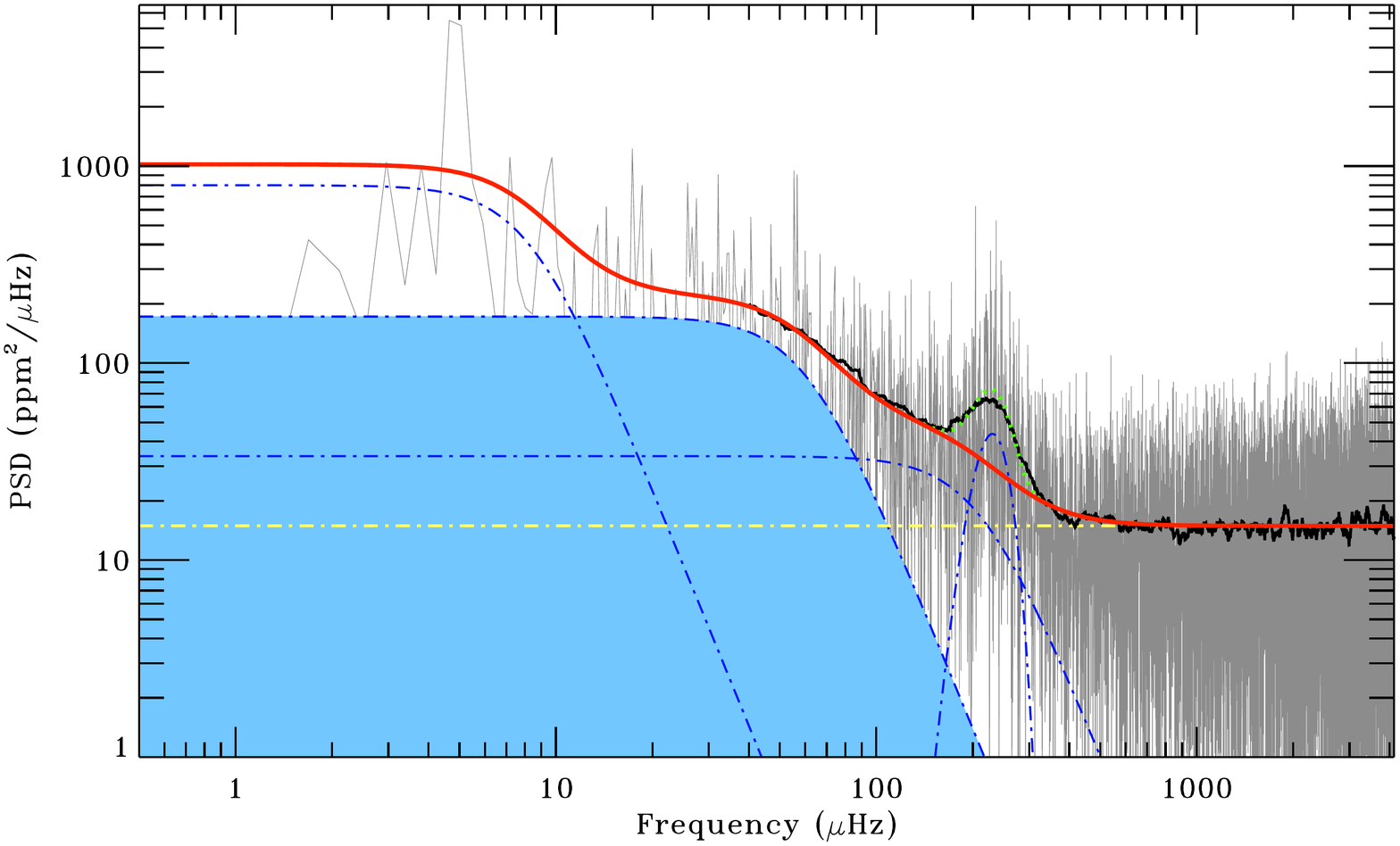}}\hfill
  \subfigure{%
  \includegraphics[width=.48\textwidth,trim={0.5cm 1.75cm 0.5cm 1.5cm},clip]{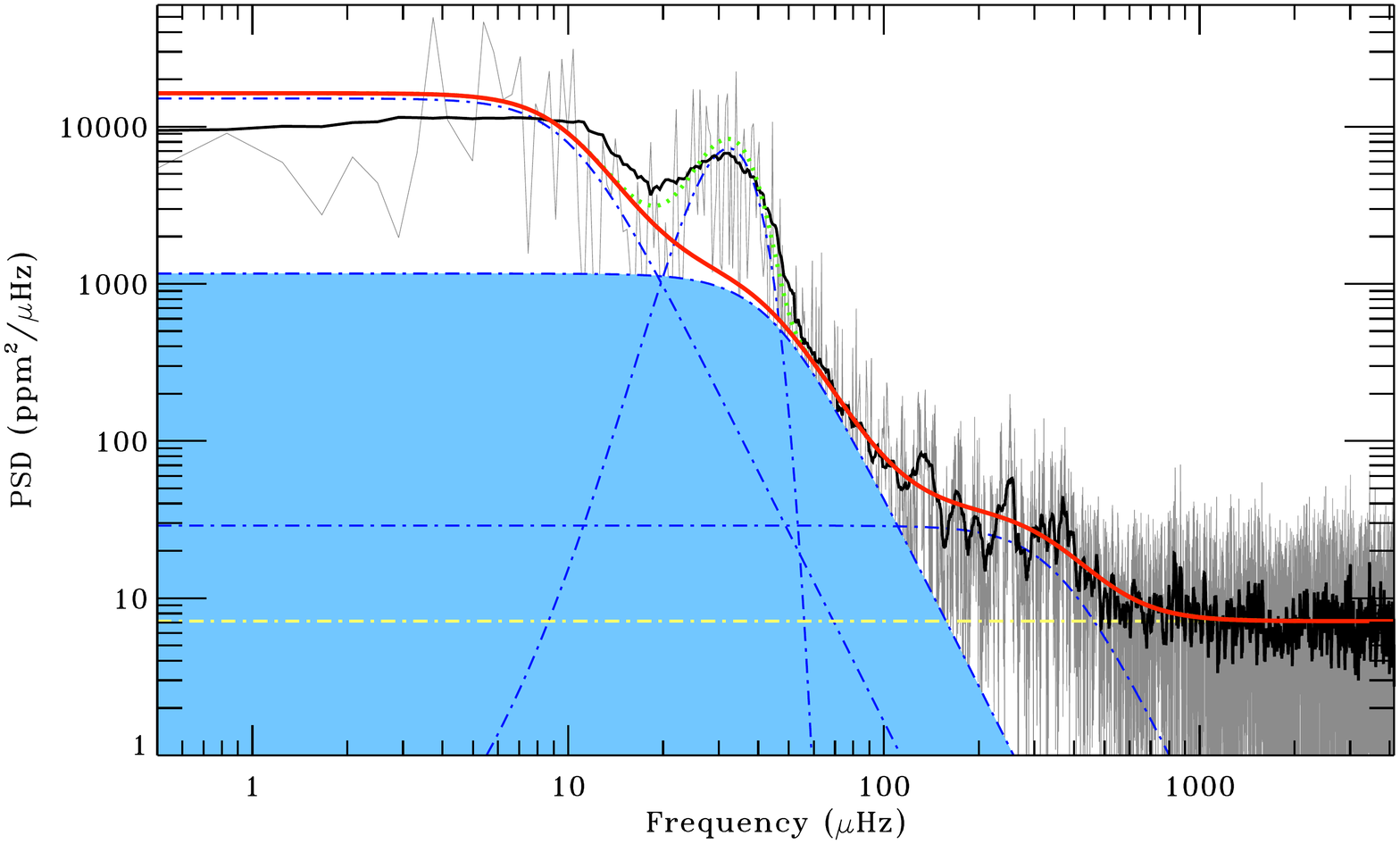}}
  \caption{Power spectral density (PSD) of HD~212771 (left panel) and HD~203949 (right panel). The PSD is shown in gray (with a heavily smoothed version in black). The solid red curve is a fit to the background, consisting of three Harvey-like profiles (blue dot-dashed curves) plus white noise (yellow horizontal dot-dashed line). A global fit to the oscillation power excess (blue dot-dashed Gaussian curve) and the background is visible at $\sim230\:{\rm \mu Hz}$ (HD~212771) and $\sim30\:{\rm \mu Hz}$ (HD~203949) as a dotted green curve.}\label{fig:psds}
\end{figure*}

\begin{figure*}[!t]
\centering
  \subfigure{%
  \includegraphics[width=.48\textwidth,trim={2cm 2.75cm 6cm 13.25cm},clip]{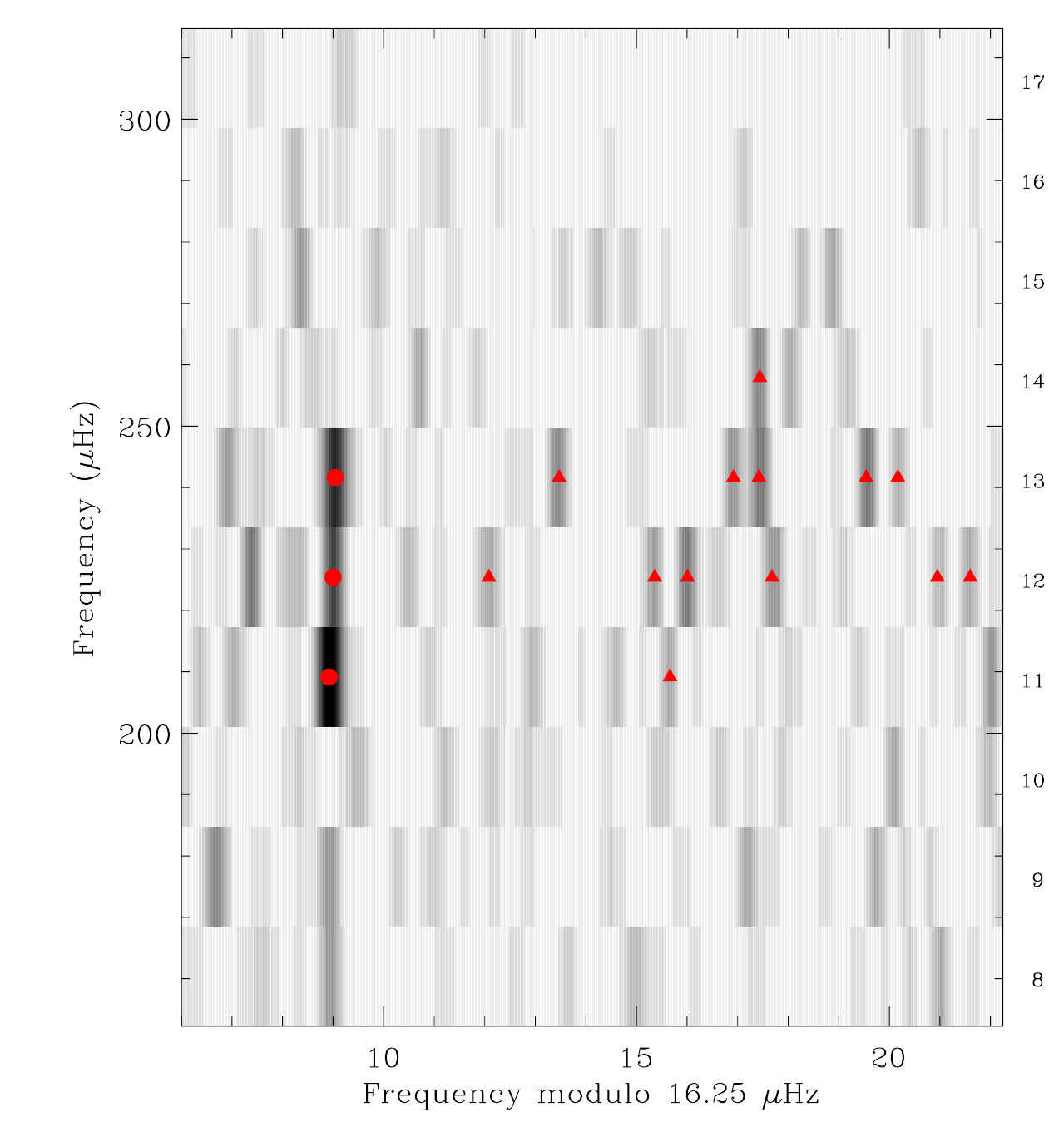}}\hfill
  \subfigure{%
  \includegraphics[width=.48\textwidth,trim={2cm 2.75cm 6cm 13.25cm},clip]{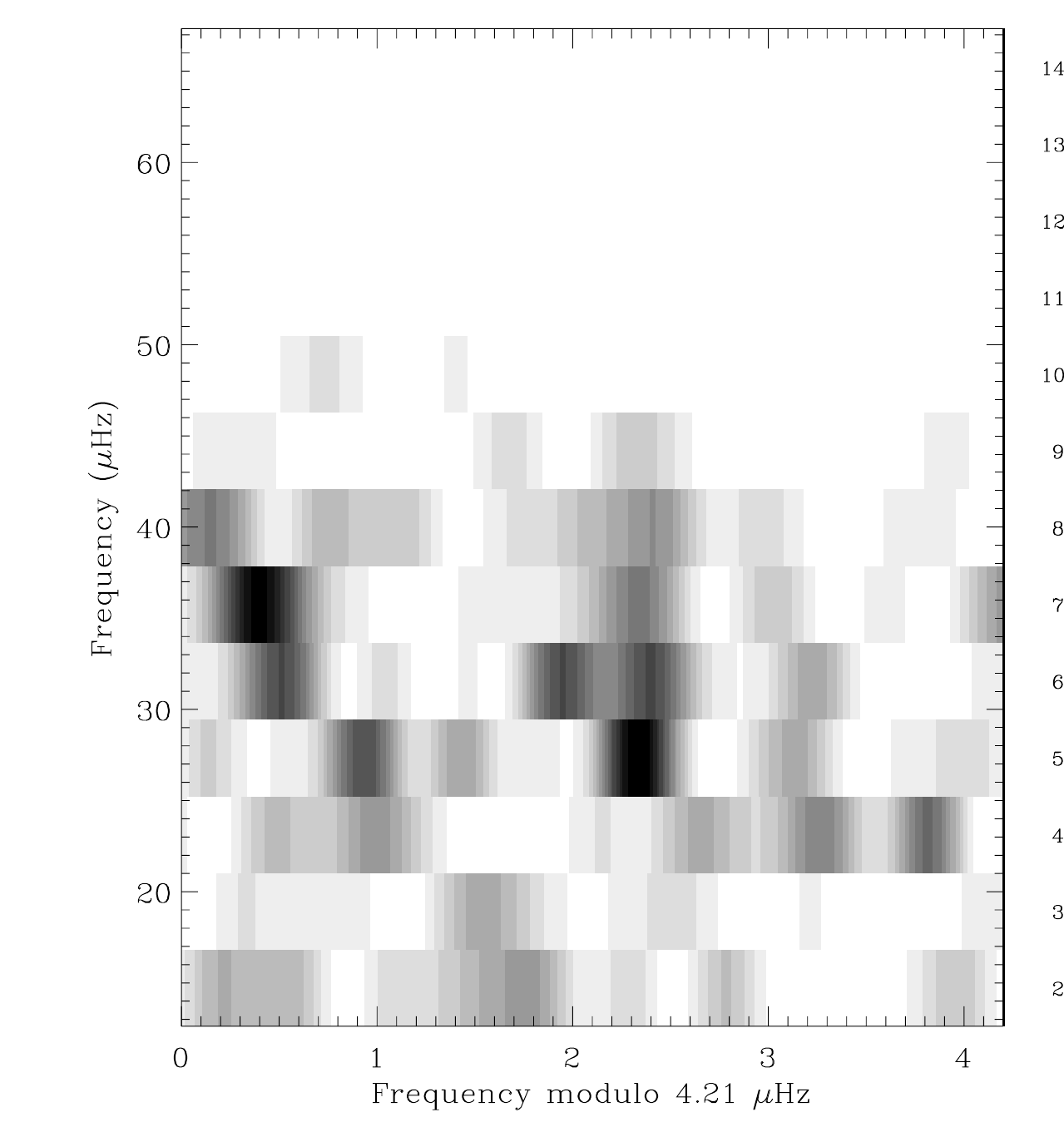}}
  \caption{Grayscale \'echelle diagram of the smoothed PSD of HD~212771 (left panel) and HD~203949 (right panel). The pressure radial order, $n_{\rm p}$, is indicated along the right $y$-axis. Identified individual modes for HD~212771 (see Fig.~\ref{fig:l1_hd212771} and Table \ref{tab:mixed}) are marked with red circles ($\ell = 0$, radial modes) and red triangles ($\ell = 1$, dipole modes). A proxy for $\Delta\nu$ of $4.21\:{\rm \mu Hz}$ (within errors of the quoted value in Table \ref{tab:HD203949}) is used in the right panel to enhance the vertical alignment of the ridges.}\label{fig:echelle}
\end{figure*}

Figure \ref{fig:psds} shows the power spectra of HD~212771 (left panel) and HD~203949 (right panel) computed based on the TASOC light curves. These reveal clear power excesses due to solar-like oscillations at $\sim230\:{\rm \mu Hz}$ and $\sim30\:{\rm \mu Hz}$, respectively. Figure \ref{fig:echelle} shows the \'echelle diagrams of the smoothed power spectra of HD~212771 (left panel) and HD~203949 (right panel).

\subsection{Global Oscillation Parameters}\label{sec:globalseismo}

The large frequency separation, $\Delta\nu$, and frequency of maximum oscillation amplitude, $\nu_{\rm max}$, were measured based on the analysis of the above power spectra. A range of well-tested and complementary automated methods were used in the analysis \citep{Huber09,Huber11,MosserAppourchaux09,Mathur10,Mosser11,DIAMONDS,Corsaro15,DaviesMiglio16,CampanteChapter,Yu18}, which have previously been extensively applied to data from \textit{Kepler}/K2 \citep[e.g.,][]{Hekker11,Verner11}. Returned values were subject to a preliminary step which involved the rejection of outliers following Peirce's criterion \citep{Peirce,Gould}. For each star, we finally adopted the values of $\Delta\nu$ and $\nu_{\rm max}$ corresponding to the smallest normalized rms deviation about the median, considering both parameters simultaneously (i.e., both parameters originate from the same source/method). Uncertainties were recalculated by adding in quadrature the corresponding formal uncertainty and the standard deviation of the parameter estimates returned by all methods. Consolidated values for $\Delta\nu$ and $\nu_{\rm max}$ are given in Tables \ref{tab:HD212771} and \ref{tab:HD203949}.

\subsection{HD~212771: Asymptotic Mixed-Mode Pattern and Rotation}\label{sec:seismo_hd212771}

Mixed modes in HD~212771 were analyzed following the method of \citet{Mosser15}, which revealed the signature of the period spacing, $\Delta\Pi_1$. Its value, computed as in \citet{Vrard16}, is $\Delta\Pi_1 = 84.3\pm1.6\:{\rm s}$. A fit of the mixed-mode pattern provides a more refined value of the period spacing, $\Delta\Pi_1 = 85.3\pm0.3\:{\rm s}$, and a coupling factor $q = 0.19\pm0.03$. Such values are in agreement with the general trends found in \textit{Kepler} data for stars on the red-giant branch \citep{Mosser17}. This is supported by the star's location in a $\Delta\Pi_1$ -- $\Delta\nu$ diagram \citep[see fig.~1 of][]{Mosser14}. We thus reclassify HD~212771 as a low-luminosity red-giant branch (LLRGB) star based on asteroseismology.

\begin{figure}[!t]
    \centering
    \includegraphics[width=\linewidth,trim=1.5cm 8.2cm 0.8cm 8.8cm,clip]{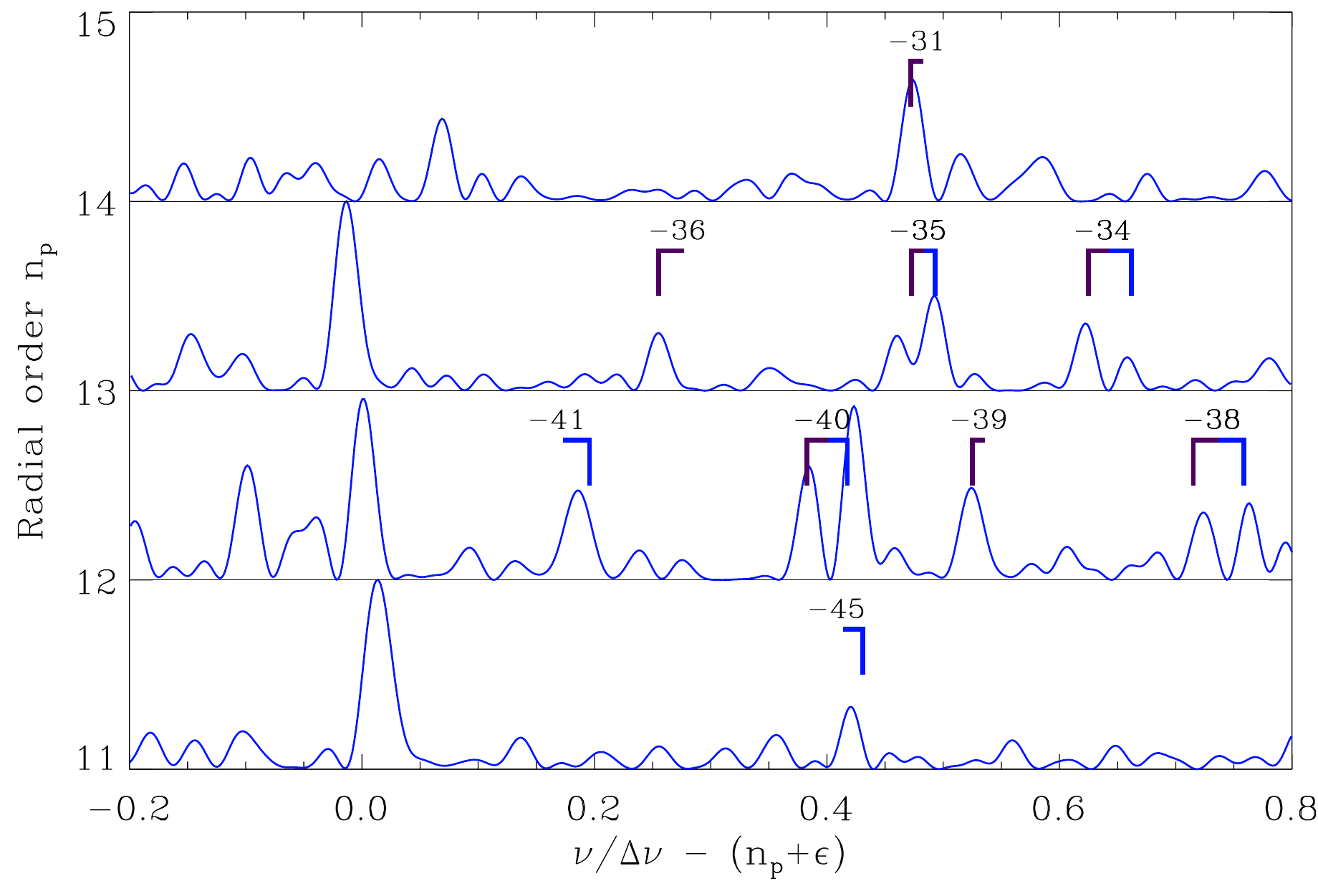}
    \caption{Mixed-mode pattern of HD~212771. The PSD along each pressure radial order, $n_{\rm p}$, is shown as a function of reduced frequency, $\nu/\Delta\nu - (n_{\rm p} + \epsilon)$, where $\epsilon$ is a phase shift sensitive to the properties of the near-surface layers of the star. The three prominent radial modes have a reduced frequency close to 0 (modulated by acoustic glitches). The mixed-mode orders, $n_{\rm m}$, are indicated, with color coding the azimuthal order ($m = -1$ in dark purple and $m = 1$ in blue; see Table \ref{tab:mixed}).}
    \label{fig:l1_hd212771}
\end{figure}

Since the mixed-mode pattern also revealed rotational multiplets, we next performed an analysis of the rotational splittings of dipole mixed modes (based on a power spectrum oversampled by a factor of 4). In a preliminary step, rotational splittings were identified using an asymptotic mixed-mode pattern modulated by a core rotation rate of $400\:{\rm nHz}$. Comparison of the spectrum with the asymptotic fit revealed 13 mixed modes with a height-to-background ratio larger than five, among which 8 are forming rotational doublets, corresponding to the mixed-mode orders $-40$, $-38$, $-35$, and $-34$ (see Fig.~\ref{fig:l1_hd212771} and Table \ref{tab:mixed}). The remaining components were identified as being $|m|=1$ modes. The absence of any significant dipole mode with azimuthal order $m=0$ is in favor of a star seen edge-on. As shown by \citet{Kamiaka18}, deriving a reliable and precise value of the stellar inclination is difficult when the height-to-background ratio of the modes is small. Based on the observed height-to-background ratio of the rotational doublets, an inclination angle larger than $75^\circ$ is to be expected, consistent with a potentially aligned transiting system.

To derive the mean core rotation, rotational splittings were expressed as a function of $\zeta$, which describes the relative contribution of the inner radiative region to the mode inertia \citep{Goupil13,Deheuvels14,Mosser18}. This method allowed us to derive the individual splitting of each component of the multiplet, with the total rotational splitting between the $m=-1$ and $m=+1$ components being split according to the respective $\zeta$ coefficients of each mode. We further assumed that the uncertainties of the unresolved dipole mixed modes cannot be less than half the frequency resolution. The method then returns a nominal, albeit imprecise, mean core rotation of $\delta\nu_{\rm rot} = 354\pm151\:{\rm nHz}$.

\begin{table}[!t]
\begin{center}
\caption{Low-Degree Oscillation Pattern of HD~212771}\label{tab:mixed}
\begin{tabular}{rrrrcccr}
\hline
$n_{\rm p}$ & $\ell$ & $m$ & $n_{\rm m}$ & $\zeta$ & $\nu_{\rm as}$ & $\nu_{\rm obs}$ & $h$ \\
 & & & & & ($\mu$Hz) & ($\mu$Hz) & \\
\hline
11 &  0 & 0 & \nodata & \nodata &  203.34 &  203.92 &  20.7 \\
11 &  1 &    1 &  $-$45 & 0.718 &  210.84 &  210.67 &   8.1 \\
12 &  0 & 0 & \nodata & \nodata &  219.73 &  220.27 &  14.8 \\
12 &  1 &    1 &  $-$41 & 0.934 &  223.49 &  223.34 &   6.2 \\
12 &  1 & $-$1 &  $-$40 & 0.775 &  226.58 &  226.62 &   8.2 \\
12 &  1 &    1 &  $-$40 & 0.680 &  227.16 &  227.27 &  12.1 \\
12 &  1 & $-$1 &  $-$39 & 0.441 &  228.94 &  228.94 &   7.6 \\
12 &  1 & $-$1 &  $-$38 & 0.884 &  232.08 &  232.21 &   5.9 \\
12 &  1 &    1 &  $-$38 & 0.910 &  232.80 &  232.86 &   5.8 \\
13 &  0 & 0 & \nodata & \nodata &  236.21 &  236.55 &  23.9 \\
13 &  1 & $-$1 &  $-$36 & 0.901 &  241.01 &  240.98 &  10.1 \\
13 &  1 & $-$1 &  $-$35 & 0.445 &  244.61 &  244.43 &   7.5 \\
13 &  1 &    1 &  $-$35 & 0.394 &  244.95 &  244.93 &  11.7 \\
13 &  1 & $-$1 &  $-$34 & 0.724 &  247.13 &  247.05 &   9.9 \\
13 &  1 &    1 &  $-$34 & 0.802 &  247.74 &  247.68 &   5.1 \\
14 &  1 & $-$1 &  $-$31 & 0.444 &  261.14 &  261.20 &  13.4 \\
\hline
\end{tabular}
\end{center}
\tablecomments{Each mode is labeled according to its pressure radial order, $n_{\rm p}$, degree, $\ell$, and azimuthal order, $m$; mixed modes are further characterized by their mixed-mode order, $n_{\rm m}$, and $\zeta$ coefficient. The asymptotic guess frequency, $\nu_{\rm as}$, is given, as well as the observed frequency, $\nu_{\rm obs}$, and height-to-background ratio, $h$. The Doppler shift of the observed frequencies due to the line-of-sight motion \citep{Davies14} is significant for both stars in this study (no correction has been applied here). We note, however, that this has a negligible effect on the analysis performed in Sect.~\ref{sec:globalseismo}.}
\end{table}

\subsection{HD~203949: What Causes the Second Power Excess?}\label{sec:seismo_hd203949}

In order to properly fit the background PSD of HD~203949, three Harvey-like profiles were required, as determined from a Bayesian model comparison using \texttt{DIAMONDS} \citep{DIAMONDS}. One of the profiles in this preferred model has, however, a timescale and amplitude that do not conform with expectations based on the measured $\nu_{\rm max}$ \citep[e.g.,][]{Kallinger14}, with a ``knee''\footnote{In the interest of reproducibility, we provide the fitted parameters of the three Harvey-like profiles \citep[for the adopted functional form and a definition of the parameters $\{a_i,b_i\}$, see eq.~4 of][]{Corsaro2017}. For HD~203949, one has $\{a_1,a_2,a_3\}=\{414^{+44}_{-45},239^{+44}_{-62},106^{+4}_{-4}\}\:{\rm ppm}$ and $\{b_1,b_2,b_3\}=\{10.2^{+1.4}_{-1.9},44.2^{+7.7}_{-8.9},352^{+16}_{-15}\}\:{\rm \mu Hz}$. For HD~212771, one has $\{a_1,a_2,a_3\}=\{85.6^{+10.6}_{-11.3},107^{+8}_{-7},88.7^{+8.0}_{-7.1}\}\:{\rm ppm}$ and $\{b_1,b_2,b_3\}=\{8.25^{+1.39}_{-1.74},60.1^{+16.7}_{-9.0},210^{+37}_{-20}\}\:{\rm \mu Hz}$.} at $\sim300\:{\rm \mu Hz}$ (see right panel of Fig.~\ref{fig:psds}). We suspect that this second power hump is caused by jitter in the TESS data that remains after applying the KASOC filter, which is particularly pronounced during the 1347 -- 1350-day period of poor spacecraft tracking (see right panel of Fig.~\ref{fig:LCs}). We note that the power hump is also evident from the simple aperture photometry (SAP) delivered by the TESS Science Processing Operations Center \citep[SPOC;][]{jenkins2016}, while it is largely removed in the co-trended Presearch Data Conditioning SAP (PDCSAP) --- the oscillation signal in the PDCSAP data is, however, of lower quality than the one present in the TASOC data.

Support for the hypothesis of jitter causing the second power hump comes from the \texttt{MOM$\_$CENTR2} data delivered by SPOC, which gives the flux-centroid along rows on the CCD. The PSD of this centroid time series shows a clear excess at $\sim300\:{\rm \mu Hz}$. Jitter at this frequency would cause a variation in the flux from inter/intra pixel sensitivity and from flux exiting/entering the aperture. We found that adopting a larger aperture than the one set by the TASOC pipeline made the power hump disappear, but at the cost of a degraded oscillation signal. Furthermore, the hypothesis that this feature is instrumental in nature is reinforced by noting that the power hump nearly vanishes if the 1347 -- 1350-day data are omitted from the PSD calculation, and that the short-cadence light curve for the nearby star TIC~129679884 shows the same effect.

Since the power hump is well separated in frequency from the power excess due to oscillations, we could account for it in the fitted background without affecting the analysis of the oscillations.

\section{Estimation of Fundamental Stellar Properties}\label{sec:properties}

Fundamental stellar properties can be estimated by comparing global asteroseismic parameters and complementary spectroscopic and astrometric data to the outputs of stellar evolutionary models. We used a number of independent grid-based pipelines in this work \citep{Stello09,Basu10,Basu12,Metcalfe10,Gai11,PARAM1,PARAM2,VSA15,Yildiz16,Serenelli17}, whereby observables are matched to well-sampled grids of stellar evolutionary tracks. The diversity of grids and optimization procedures employed implicitly account for the impact of using different stellar models --- covering a range of input physics --- and analysis methodologies on the final estimates. The adopted set of observables consists of \{$\Delta\nu$, $\nu_{\rm max}$, $[{\rm Fe}/{\rm H}]$, $T_{\rm eff}$, $L_\ast$\}. Given the negligible $\alpha$ enhancement (see Tables \ref{tab:HD212771} and \ref{tab:HD203949}), we have neglected its effect.

We provide consolidated values from grid-based modeling for the stellar mass, $M_\ast$, radius, $R_\ast$, mean density, $\rho_\ast$, surface gravity, $\log g$, and age, $t$, in Tables \ref{tab:HD212771} and \ref{tab:HD203949}. To properly account for systematics, values returned by the several pipelines were subject to the same procedure as described in Sect.~\ref{sec:globalseismo} (i.e., single source/method), except that no preliminary outlier rejection step has now been applied.

The properties estimated for HD~212771 in this work are consistent with those estimated by \citet{Campante17} based on K2 asteroseismology. As noted by those authors, the derived asteroseismic mass places HD~212771 just within the \textit{retired A star} category\footnote{RV planet surveys rely on evolved stars for a sample of intermediate-mass stars ($M_\ast \gtrsim 1.5\,{\rm M}_\odot$; so-called retired A stars), which are more amenable to RV observations than their main-sequence counterparts.}, being significantly larger than the value reported in the discovery paper \citep[see also][]{North17}.

Regarding HD~203949, we provide fundamental properties assuming that the star is either on the red-giant branch (RGB) or in the clump, deferring a discussion of its evolutionary state until Sect.~\ref{sec:evolstate}. We note the large discrepancy between both asteroseismic masses derived in this work and the mass quoted in the discovery paper \citep[$M_\ast = 2.1 \pm 0.1\,{\rm M}_\odot$;][]{Jones14}. \citet{Jones14} identified HD~203949 as a post-RGB star. Their large mass determination can, to a large extent, be ascribed to the surface gravity adopted, $\log g = 2.94 \pm 0.20$, consistent with stars in the secondary clump and hence masses $\gtrsim 2\,{\rm M}_\odot$. Under the assumption that HD~203949 is in the clump, the lower seismic gravity (see Table \ref{tab:HD203949}) is consistent with that of a typical red-clump star, ruling out a large mass. We stress here that asteroseismology can be used to accurately and robustly determine surface gravities for red giants, with systematic offsets of only a few percent \citep{pinsonneault:2018}.

This large mass discrepancy calls for a revision of both the planet's semimajor axis and minimum mass. By assuming an RV semi-amplitude of $178.1 \pm 10.0\:{\rm m\,s^{-1}}$ and an orbital period of $184.2 \pm 0.5$ days \citep{Jones14}, we find that in the RGB scenario, $a = 0.68 \pm 0.03\:{\rm au}$ and $M_{\rm p} \sin i = 5.7 \pm 0.6\,M_{\rm J}$, whereas in the clump scenario, $a = 0.63 \pm 0.04\:{\rm au}$ and $M_{\rm p} \sin i = 5.0 \pm 0.6\,M_{\rm J}$. In both cases, the parameters were derived assuming a circular orbit, in line with the observed eccentricity. The revision of the planet's properties thus implies a change $>30\,\%$ in its estimated mass.


\section{Discussion}\label{sec:discuss}

\subsection{Asteroseismic Performance: TESS vs.~\textit{Kepler}/K2}\label{sec:perf}
HD~212771 was observed by K2 in short cadence \citep{Campante17,North17}, which enabled its asteroseismic investigation. Here, we compare the asteroseismic performances of K2 and TESS by assessing the ratio of the observed maximum oscillation amplitudes for this star, i.e., $A_{\rm max}^{\rm TESS}/A_{\rm max}^{\rm K2}$.

The absolute calibration of the oscillation amplitudes depends on the instrument's bandpass. TESS has a redder bandpass than \textit{Kepler}/K2, meaning observed amplitudes are expected to be lower in the TESS data by a factor of $\sim0.85$ \citep{Campante16}. Based on a black body approximation, M.~N.~Lund (2019, in prep.) finds this factor to be slightly lower, i.e., $\sim0.83$--$0.84$ on average within the $T_{\rm eff}$ range considered in that study. We measured the maximum oscillation amplitude per radial mode, $A_{\rm max}$, following the method introduced by \citet{Kjeldsen05,Kjeldsen08}, which involves determining the peak of the heavily smoothed, background-corrected amplitude oscillation envelope having accounted for the (bandpass-dependent) effective number of modes per radial order\footnote{We used the same K2 light curve as in \citet{Campante17} in our analysis.}. This yielded $A_{\rm max}^{\rm TESS} = 12.8 \pm 2.3\:{\rm ppm}$ and $A_{\rm max}^{\rm K2} = 17.1 \pm 0.9\:{\rm ppm}$, resulting in a ratio $A_{\rm max}^{\rm TESS}/A_{\rm max}^{\rm K2} = 0.75 \pm 0.14$, consistent with the expected ratio.

We caution the reader that the estimated $A_{\rm max}^{\rm TESS}/A_{\rm max}^{\rm K2}$ is prone to unaccounted biases due to the stochastic nature of the oscillations \citep[e.g.,][]{Arentoft19}, especially when considering the short time coverage compared to the lifetime of the modes as well as the non-contemporaneity of the TESS and K2 datasets. Moreover, the absolute values of $A_{\rm max}^{\rm TESS}$ and $A_{\rm max}^{\rm K2}$, taken individually, are also subject to biases arising from the choice of background model. Their ratio, however, can be more accurately estimated if both values are computed assuming the same functional form for the background model, which has been done here. Despite the above, this preliminary, single-point estimate of $A_{\rm max}^{\rm TESS}/A_{\rm max}^{\rm K2}$ provides support for the predicted yield of solar-like oscillators using TESS's 2-minute cadence observations \citep{Schofield19}.

\subsection{On the Evolutionary State of HD~203949: RGB vs.~RC}\label{sec:evolstate}

Red-clump (RC) stars, i.e., cool He-core burning stars, occupy a confined parameter space in the $\Delta\Pi_1$ -- $\Delta\nu$ diagram around $300\:{\rm s}$ and $4.1\:{\rm \mu Hz}$ \citep{Mosser12}. Although the $\Delta\nu$ value measured for HD~203949 is consistent with it being an RC star, the low frequency resolution of the power spectrum hinders a measurement of $\Delta\Pi_1$. This in turn prevents a definitive classification of its evolutionary state from being made based on the $\Delta\Pi_1$ -- $\Delta\nu$ diagram, due to the underlying degeneracy for $\Delta\nu \lessapprox 10\:{\rm \mu Hz}$ \citep[e.g.,][]{Mosser14}. In an attempt to assess the evolutionary state of HD~203949, we have thus conducted a number of analyses, which we summarize below.

\textbf{Machine Learning Classification.} We employed the deep learning method of \citet{Hon17,Hon18}, which  efficiently classifies the evolutionary state of oscillating red giants by recognizing visual features in their asteroseismic power spectra. A test set accuracy of 93.2\,\% has been reported when applying the classifier to 27-day photometric time series \citep{Hon18}. Application of this method to the power spectrum of HD~203949 returns a probability of it being an RC star of $p \sim 0.6$, having taken into account the effect of detection bias in the training set.

Alternatively, we have made use of \texttt{Clumpiness} (J.~Kuszlewicz et al.~2019, in prep.). \texttt{Clumpiness} uses a handful of well-engineered features and a gradient boosting algorithm \citep[\texttt{xgboost};][]{Chen16} to classify stars as RGB or RC (or even as possible main-sequence stars observed in long cadence) in the time domain. These features include the median absolute deviation from the median (MAD) of the time series flux, the number of zero crossings, a measure of the stochasticity following \citet{Kedem81,Kedem82,Bae96}, the MAD of the first differences, and, to complement the time series features, the $K$-band absolute magnitude is also included, which is computed using distances from \citet{BailerJones18} and a 3D dust map from \citet{Green15}. Across a range of time series lengths, from 27 days up to 4 years, the classifier maintains an accuracy of approximately 92\,\%. Computing the features for HD~203949, the classifier returns a probability of 0.6 of it being in the RC.

\textbf{Grid-Based Modeling.} For a given set of seismic and spectroscopic observational constraints, the evolutionary state of HD~203949 can also be assessed from the results of grid-based modeling. We have thus performed two separate analyses, each assuming as prior information a specific evolutionary state, i.e., RGB or RC. We found that the probability of HD~203949 being an RC star is 75 times greater than it being an RGB star (or $p = 0.99$), as determined by the ratio of the overall posterior probabilities of both scenarios. Interpreting this in terms of a Bayes' factor provides very strong evidence in support of the RC scenario given the adopted set of seismic and spectroscopic constraints. We looked into which observational constraints are driving this result by analyzing their posterior distributions. RC stellar models reproduce very well all constraints, while RGB models cannot simultaneously fit the effective temperature and metallicity for stellar masses that are compatible with the seismic data. RGB models with $1\,{\rm M}_\odot$ are too cool for [Fe/H]=0.17 by about $200\:{\rm K}$ or, conversely, too metal rich for the observed effective temperature by about 0.35 dex. The temperature difference between the RC and the RGB at fixed $\log g$ is smaller for tracks of larger masses, since the effective temperature of the clump does not change while that of the RGB gets higher. Our grid-based modeling for the RGB scenario reflects this, yielding a higher mass, around $1.2\,{\rm M}_\odot$. This higher mass is obtained at the expense of a posterior $\nu_{\rm max}$ higher than, and in tension with, the observed value. Finally, it is worth mentioning that this conclusion is robust against the temperature scale defined by the choice of mixing length, $\alpha_{\rm MLT}$, in the stellar models. We tested models with $\alpha_{\rm MLT}$ ranging from 1.8 (solar-calibrated value with an Eddington atmosphere) to 2.1 (solar-calibrated value with a Krishna Swamy atmosphere), with almost no impact on the Bayes' factor, which varied from 75 for $\alpha_{\rm MLT} = 1.8$ down to 70 for $\alpha_{\rm MLT} = 2.1$.

We have also employed the Bayesian inference method of \citet{Stock18}. This method compares the position of a star in the Hertzsprung--Russell diagram with those of the latest \texttt{PARSEC} evolutionary models \citep{PARSEC}. The spectroscopically-determined metallicity, $B-V$ color, and astrometry-based luminosity \citep{Arenou99}, computed from the adjusted \textit{Gaia} parallax in Table \ref{tab:HD203949}, were used as constraints. Moreover, the initial mass function and the evolutionary timescale at each model position were used as priors in the Bayesian inference. The outcomes are probability density functions for the stellar parameters as well as probabilities of the star being either on the RGB or the horizontal branch. The method was carefully tested by \citet{Stock18} against reference samples with accurate stellar parameters determined using different methods, and was found to deliver very reliable results. In particular, its reliability was tested against a sample of evolved stars with evolutionary states determined from asteroseismology, resulting in an accuracy of 86\,\%. Application of this method returns a probability of HD~203949 being an RC star of $p = 0.93$.

\textbf{Asymptotic Acoustic-Mode Offset.} \citet{Kallinger12} found an empirical relation between the asymptotic offset, $\epsilon_{\rm c}$, of radial modes in red giants and the evolutionary state, separating H-shell (RGB) from He-core burning stars. \citet{JCD14} provided a theoretical interpretation of this relation, which was found to derive from differences in the thermodynamic state of the convective envelope. Both works acknowledge the potential of this relation for distinguishing RGB and clump stars when faced with observations that are too short to allow such a distinction based on the determination of $\Delta\Pi_1$. We extracted frequencies\footnote{Note that these frequencies are not obtained from a full peak-bagging analysis. Nevertheless, they are still reliable and match observed peaks in the PSD.} from the PSD of HD~203949 using the multi-modal approach described in \citet{Corsaro19}. The value of $\epsilon_{\rm c}$ was then constrained from an \'echelle diagram and found to be $1.24 \pm 0.05$. The quoted uncertainty corresponds to the width of the expected $\ell\!=\!0$ ridge in the \'echelle diagram and reflects the lack of resolving power to properly disentangle radial modes from adjacent quadrupole modes. We note, however, that this uncertainty is not consistent with the one in $\Delta\nu$ and should thus be considered as a lower limit. An uncertainty of 0.05 in $\epsilon_{\rm c}$ translates into a relative uncertainty in $\Delta\nu$ of about $0.05/n_{\rm max} < 1\,\%$ \citep[e.g.,][]{Mosser13}, with $n_{\rm max}$ the radial order at $\nu_{\rm max}$. Table \ref{tab:HD203949} nevertheless quotes a relative uncertainty in $\Delta\nu$ of $3.4\,\%$. Once the uncertainty in $\epsilon_{\rm c}$ has been calibrated, the measured value of $\epsilon_{\rm c} = 1.2 \pm 0.2$ then allows for both evolutionary states (within $1.5\,\sigma$) in the top panel of fig.~4 of \citet{Kallinger12}.

\textbf{Spectroscopic evolutionary state.}
We made an attempt at inferring the spectroscopic evolutionary state of HD~203949 as described in \citet{Holtzman18}. The basic idea behind this approach is to use a ridgeline in the $T_{\rm eff}$ -- $\log g$ plane that is a function of metallicity, and supplement this with a measurement of the surface [C/N] ratio, since the latter is expected (and observed) to further separate the RGB and RC. This approach was devised to separate RGB and RC stars in the (asteroseismic) APOKASC sample \citep{pinsonneault:2018} and has an accuracy of approximately 95\,\%. In the absence of a [C/N] measurement, we estimated the range of possible values from the stellar mass, taking as reference the APOKASC sample. This led to the star being most likely in the RC. However, we also have to account for the fact that the above relations are defined in the APOKASC sample only and that there could be a systematic offset between the $T_{\rm eff}$ and/or metallicity scales. To test this, we computed the photometric temperature by means of the infrared flux method \citep{GH&B09}, leading to a temperature cooler than the spectroscopic one (at the 1.5\,$\sigma$ level). Adopting the photometric temperature, one instead arrived at the RGB classification. The issue of evolutionary state hence seems to rely sensitively on the effective temperature, given that the error on the spectroscopic $T_{\rm eff}$ is large enough to encompass both scenarios.

In summary, all but one approach give an ambiguous answer. The spectroscopic evolutionary state is unresolved due to the possibility of a systematic offset between the $T_{\rm eff}$ scales. The asymptotic acoustic-mode offset, $\epsilon_{\rm c}$, whose uncalibrated uncertainty is a poorly constrained lower limit, allows for both evolutionary states. Despite its inconclusiveness in this particular instance, machine learning classification still exhibits a high degree of accuracy, thus holding great promise for large ensemble studies with TESS (e.g., Galactic archaeology). Finally, the two applications of grid-based modeling provide very strong evidence in support of the red-clump scenario given the adopted set of seismic and spectroscopic constraints. Although the balance of evidence seems to favor the RC scenario, there are two points worth noting. First, there has been rising concern that standard RC models might suffer from important underlying systematic errors \citep[e.g.,][]{An19}, which could undermine results coming out from the grid-based modeling approach. Second, there is no direct observational evidence decisively pointing to either scenario. In light of the above, we thus refrain from providing a definitive classification of the evolutionary state of HD~203949.

\subsection{Orbital Evolution of HD~203949 b: Avoiding Engulfment at the Tip of the RGB }\label{sec:orbevol}

The history, evolution and fate of the planet orbiting HD~203949 change significantly depending on whether the star is an RGB or a red-clump star. The more straightforward scenario is that HD~203949 is in the process of ascending the RGB and will eventually engulf the orbiting planet. The alternative scenario, in which HD~203949 is in the red clump, calls for a more detailed examination. We now go on to discuss this scenario.

The variations in radius, luminosity and mass of giant-branch stars often have destructive consequences for planetary systems \citep{veras2016}. Most important for HD~203949 b is the radius variation of the host star, which could incite star-planet tides that might engulf the planet \citep{villiv2009,kunetal2011,musvil2012,adablo2013,norspi2013,valras2014,viletal2014,madetal2016,staetal2016,galetal2017,raoetal2018}. The asteroseismic stellar mass of $1.00 \pm 0.16\,{\rm M}_\odot$ (under the clump assumption) would tidally influence and probably lead to the engulfment and destruction of a Jovian planet on a 184-day orbit at the tip of the RGB.

A planet which is engulfed in the low-density atmosphere of a giant-branch star usually decays quickly enough for it to be considered destroyed. Figure 4 of \cite{macetal2018} estimates decay times of engulfed Jovian planets across the Hertzsprung--Russell diagram, and finds that the spiral-in process lasts $10^0$--$10^4$ orbits. The upper bound of this range (corresponding to about 5000 yr in our case) is much less than the timescale (about 2 Myr) in which this star's radius would exceed $a$ (0.63 au) during the RGB phase. Hence, HD~203949 b would unlikely have survived being engulfed.

Now let us assume that the planet would avoid being engulfed. In general, there are two outcomes: (1) the outward expansion of the planet's orbit due to stellar mass loss dominates over tidal effects, and the planet's final semimajor axis increases, or (2) tidal effects dominate over mass loss, but only for a short enough time to prevent engulfment, leading to a decrease in the final semimajor axis. Outcome (2) is expected to be rare because the engulfment timescale is so small. Nevertheless, this outcome may explain the current orbit of HD~203949 b under the clump assumption.

We explored this possibility by performing numerical simulations of star-planet tides, with the intention of providing rough estimates\footnote{Not considered here are the effects of evaporation of the planet's atmosphere due to the RGB stellar luminosity.}. We used four different stellar tracks with different values of the Reimers' mass-loss coefficient, $\eta$, metallicity and atmospheric type (Krishna Swamy and Eddington, which lead to different model $T_{\rm eff}$ scales on the RGB and hence different stellar radii), which fit the currently measured stellar observables. In all cases, a planetary semimajor axis corresponding to a 184-day period (0.63 au) is well within the maximal radial extent of the star, which is attained at the tip of the RGB and ranges from 0.85--0.99 au across the four tracks. Therefore, outcome (2) from above would apply to this system.

The extent to which the planet would be dragged inward changes depending on the details of the tidal formalism adopted. We used a basic formulation of dynamical tides from \cite{zahn1977}, as implemented in \cite{viletal2014}, by (i) including frictional forces from the stellar envelope, (ii) adopting velocity and density prescriptions from eqs.~(53) and (54) of \cite{veretal2015}, (iii) assuming zero eccentricity throughout the simulation, (iv) assuming a planetary radius of $1.0\,R_{\rm J}$, and (v) assuming adiabatic stellar mass loss, which is a robust approximation for this system \citep{veretal2011}.

In order for the planet to achieve an orbit with a semimajor axis of 0.6--1.0 au, we hence find that the main-sequence semimajor axis of the planet would have resided within an extremely narrow range (an interval much smaller than $10^{-2}$ au) centered on a specific value within the interval 3.1--3.5 au (which is set by the stellar model adopted and details of the tidal prescription). This result makes sense in the context of, for example, figs.~1, 4 and 6 of \cite{viletal2014}.

A different, but also viable explanation for the current 0.63 au orbit would be for the planet to have been gravitationally scattered into its current position after the host star had reached the tip of the RGB\footnote{Some tidal circularization might have followed the scattering event, as scattering alone usually excites rather than damps orbital eccentricity.}. Although RGB mass loss might have triggered the instability leading to this scenario \citep{debsig2002,veretal2013}, more recent suites of simulations of multiple giant-planet systems demonstrate that post-mass-loss scattering events --- at least for single stars\footnote{\cite{Jones14} did detect a long-term linear trend in the RV residuals, which might be attributed to the presence of a distant stellar companion. However, no constraints were placed on the mass nor orbital period of this putative companion.} --- are usually delayed until the white dwarf phase \citep{musetal2014,musetal2018,vergae2015,veretal2016,veretal2018}. Increasing the feasibility of gravitational scattering is that those studies adopted more massive stars than HD~203949, and hence would harbor shorter giant-branch lifetimes in which scattering could occur.

\section{Outlook}\label{sec:conclusions}

Characterization of host stars is a critical component of understanding their planets. For example, the radius of the star is required to estimate the radius of the planet from transit observations, and the luminosity and effective temperature of the star are crucial ingredients for determining the incident flux received by the planet and the extent of the Habitable Zone \citep{kane2014,kane2016}. For known systems observed with TESS, the combination of precision photometry with asteroseismology will aid in the assessment of potential transit events for RV planets \citep{dalba2019}. Dynamical studies of planetary systems require detailed knowledge of the stellar properties, such as the stellar mass \citep{menou2003}. Furthermore, the evolution of orbits as stars move off the main sequence depends on the stellar mass and radius, as these relate to the mass loss relative to the progenitor \citep{damiani2018}. Additionally, the angular size of the host star will be invaluable information when considering known systems as potential direct imaging targets \citep{kane2018}. Finally, accurate stellar radii for evolved stars will greatly improve transit probability estimates. Transit probabilities for evolved stars tend to have the largest values, since they scale linearly with stellar radius \citep{kane2010}. The asteroseismology techniques described here are thus an important component of overall planetary system characterization.

\acknowledgments

This paper includes data collected by the TESS mission. Funding for the TESS mission is provided by the NASA Explorer Program. Funding for the TESS Asteroseismic Science Operations Center at Aarhus University is provided by ESA PRODEX (PEA 4000119301) and Stellar Astrophysics Centre (SAC), funded by the Danish National Research Foundation (Grant agreement No.: DNRF106). The project leading to this publication has received funding from the European Union's Horizon 2020 research and innovation programme under the Marie Sk\l{}odowska-Curie grant agreement No.~792848 (PULSATION). This work was supported by FCT/MCTES through national funds (UID/FIS/04434/2019). This work was supported by FCT through national funds (PTDC/FIS-AST/30389/2017, PTDC/FIS-AST/28953/2017 \&\linebreak PTDC/FIS-AST/32113/2017) and by FEDER through COMPETE2020 (POCI-01-0145-FEDER-030389, POCI-01-0145-FEDER-028953 \& POCI-01-0145-FEDER-032113). This research was supported in part by the National Science Foundation under Grant No.~NSF PHY-1748958 through the Kavli Institute for Theoretical Physics program ``Better Stars, Better Planets''. The research leading to the presented results has received funding from the European Research Council under the European Community's Seventh Framework Programme (FP7/2007-2013)/ERC grant agreement No.~338251 (StellarAges). E.C.~is funded by the European Union's Horizon 2020 research and innovation programme under the Marie Sk\l{}odowska-Curie grant agreement No.~664931. M.N.L.~acknowledges support from the ESA PRODEX programme. B.M.~and R.A.G.~acknowledge the support received from CNES through the PLATO grants. A.S.~is partially supported by grants ESP2017-82674-R (Spanish Government) and 2017-SGR-1131 (Generalitat de Catalunya). D.V.~gratefully acknowledges the support of the STFC via an Ernest Rutherford Fellowship (grant ST/P003850/1). V.A.~and S.G.S.~acknowledge support from FCT through Investigador FCT contracts No.~IF/00650/2015/CP1273/CT0001 and No.~IF/00028/2014/CP1215/CT0002, respectively. S.B.~acknowledges NSF grant AST-1514676 and NASA grant NNX16AI09G. S.M.~acknowledges support from the Spanish Ministry through the Ram\'on y Cajal fellowship No.~RYC-2015-17697. M.B.N.~acknowledges support from NYUAD Institute grant G1502. S.R.~acknowledges support from the DFG priority program SPP 1992 ``Exploring the Diversity of Extrasolar Planets (RE 2694/5-1)''. M.Y., Z.\c{C}.O., and S.\"O.~acknowledge the Scientific and Technological Research Council of Turkey (T\"UB\.ITAK:118F352). D.H.~acknowledges support by the National Aeronautics and Space Administration (80NSSC18K1585, 80NSSC19K0379) awarded through the TESS Guest Investigator Program. M.S.C.~is supported in the form of a work contract funded by FCT (CEECIND/02619/2017). H.K.~acknowledges support from the European Social Fund via the Lithuanian Science Council grant  No.~09.3.3-LMT-K-712-01-0103.

%

\vspace{5mm}
\facilities{TESS, \textit{Gaia}, \textit{Kepler}(K2), VLT:Kueyen(UVES), Max Planck:2.2m(FEROS)}


\software{TASOC photometry pipeline \citep[\url{https://github.com/tasoc};][]{TDAdr},
\texttt{DIAMONDS} \citep[\url{https://github.com/EnricoCorsaro/DIAMONDS};][]{DIAMONDS}
}

\bibliography{references}



\end{document}